\newif\ifonecolumn
\onecolumntrue 
\onecolumnfalse
\ifonecolumn
  \documentclass[journal,draftcls,onecolumn, 12pt]{IEEEtran}
\else
  \documentclass[journal]{IEEEtran}
\fi
\usepackage[T1]{fontenc}

\usepackage{amsmath}
\usepackage{cite}
\usepackage{array}
\usepackage{stfloats}
\usepackage{diagbox}
\usepackage{url}
\usepackage{amsmath,amssymb}
\usepackage{epsfig,fancyhdr}
\usepackage{CJKutf8}
\usepackage{times}
\usepackage{graphicx}
\usepackage{bm}

\newcommand{\nn}{\nonumber}

\newtheorem{thm}{Theorem}

\newtheorem{defin}[thm]{Definition}
\newtheorem{lemma}[thm]{Lemma}
\newtheorem{crly}[thm]{Corollary}
\newtheorem{eg}{Example}

\begin{document}
\title{How to Construct Mutually Orthogonal Complementary Sets with Non-Power-of-Two Lengths?}

\author{Shing-Wei~Wu \begin{CJK}{UTF8}{bsmi} (吳星蔚)\end{CJK}, Chao-Yu~Chen \begin{CJK}{UTF8}{bsmi} (陳昭羽)\end{CJK},~\IEEEmembership{Member,~IEEE}, and Zilong~Liu \begin{CJK}{UTF8}{bsmi} (劉子龍)\end{CJK},~\IEEEmembership{Member,~IEEE}
\thanks{Manuscript received October 15, 2019; accepted February 14, 2020.
The work of S.-W. Wu and C.-Y. Chen was supported in part by the Ministry of Science and Technology,
Taiwan, R.O.C., under Grant MOST 107--2221--E--006--065--MY2.
}
\thanks{Chao-Yu~Chen and Shing-Wei~Wu are with the Department
of Engineering Science, National Cheng Kung University, Tainan 701, Taiwan, R.O.C.
(e-mail: \{super, n98084016\}@mail.ncku.edu.tw).}
\thanks{Zilong~Liu is with the School of Computer Science and Electronic Engineering, University of Essex, UK, (e-mail: zilong.liu@essex.ac.uk).}
\thanks{Copyright (c) 2017
IEEE. Personal use of this material is permitted. However, permission to use this material for any
other purposes must be obtained from the IEEE by sending a request to pubspermissions@ieee.org. Permission from IEEE must be obtained for all other uses, in any current or future media, including reprinting/republishing this material for advertising or promotional purposes, creating new collective works, for resale or redistribution to servers or lists, or reuse of any copyrighted component of this work in other works.}
\thanks{DOI:10.1109/TIT.2020.2980818}
}


\maketitle

\begin{abstract}

Mutually orthogonal complementary sets (MOCSs) have received significant research attention in recent years due to their wide applications in communications and radar. Existing MOCSs which are constructed based on generalized Boolean functions (GBFs) mostly have lengths of power-of-two. How to construct MOCSs with non-power-of-two lengths whilst having large set sizes is a largely open problem. With the aid of GBFs, in this paper, we present new constructions of such MOCSs and show that the maximal achievable set size is $1/2$ of the flock size of an MOCS.
\end{abstract}


\section{Introduction}
The concept of Golay complementary pair (GCP), where the sums of aperiodic autocorrelations of two sequences are zero everywhere except at the in-phase position, was introduced by Marcel J. E. Golay in his design of infrared multislit spectrometry \cite{Golay}. 
In 1972, Tseng and Liu generalized the concept to Golay complementary sets (GCSs) and mutually orthogonal complementary sets (MOCSs) \cite{Golay_sets}. Specifically, each GCS is composed of two or more sequences whose aperiodic autocorrelation sums display ``impulse-like" correlation properties as GCP; MOCSs may be viewed as a collection of GCSs in which any two distinct GCSs are mutually orthogonal in terms of their zero cross-correlation sums for all the time-shifts. In 1988, Suehiro and Hatori proposed complete complementary codes (CCC) whose set size (denoted by $M$) achieves the upper bound of MOCSs (i.e., $M\leq N$, where $N$ refers to the number of constituent sequences in an MOCS, called the flock size) \cite{N_shift}. Due to the ideal auto- and cross- correlation properties, MOCSs (and CCCs) have been adopted as a key component for next generation multicarrier code division multiple access (MC-CDMA) systems \cite{Bell_CDMA,Chen_CDMA,Chen2007book,Liu2014,Liu2015}. Other applications include optimal channel estimation in multiple-input and multiple-output (MIMO) frequency-selective channels \cite{Wang07}, MIMO radar  \cite{Li2010,Tang2014}, cell search in orthogonal frequency division multiplexing (OFDM) systems \cite{ChenICC08}, and data hiding \cite{Kojima2014}, et al.

A primary approach to construct MOCSs is based on generalized Boolean functions (GBFs). This was initiated by the landmark work of Davis and Jedwab in \cite{Golay_RM} which shows that $2^h$-ary GCPs lie in certain second-order cosets of the first-order Reed-Muller (RM) codes. Subsequently, Paterson generalized Davis and Jedwab's methods to connect $q$-ary (for even $q$) GCPs with GBFs of degree 2 \cite{Paterson_00}. In the literature, GCPs obtained on GBFs given in \cite{Golay_RM, Paterson_00} are sometimes called the Golay-Davis-Jedwab (GDJ) pairs \cite{Li_05,even-ZCP,odd-ZCP}. For GCSs, further constructions of GCSs based on GBFs of high degrees have been proposed in \cite{schmidt, Parker_03, Tellambura_ICC05, ChenAAECC06}. In \cite{MutualGCS_2008,Chen08}, connections of CCCs, MOCSs and cosets of the first-order RM codes have been proposed.

Existing complementary sequences constructed based on GBFs are mostly limited to lengths with powers of two. Recently, some research attempts have been made by Chen for GCSs with more flexible lengths \cite{Super_16,Super_172,Super_18}. GCSs with flexible lengths are of interest to practical OFDM systems with varying numbers of subcarriers. With the aid of GBFs, however, \textit{how to construct MOCSs with non-power-of-two lengths whilst having large set size} remains largely open, to the best of our knowledge. Although some MOCS constructions based on a series of sequence operations (e.g., reversals, negations, interleaving, and concatenations) have been proposed in \cite{Golay_sets}, they are unable to give rise to MOCSs with large set sizes. Moreover, these sequence operations may not be friendly for efficient hardware generations. By contrast, the sequences from GBFs have algebraic structures and hence enjoy efficient synthesis. In \cite{Das_18, Das_18_2, Das_19}, Das {\it et al.} presented methods by using paraunitary (PU) matrices to construct MOCSs and CCCs with various lengths. PU matrices are powerful tools but 1) the available MOCSs heavily depend on the existence of PU matrix kernels and 2) the generations of MOCSs require multiple blocks in serial to carry out complicated PU matrix multiplications. From these perspectives, comparison of the MOCSs based on PU matrices and that based on GBFs is not straightforward.

In this paper, we attack the aforementioned open problem by using GBFs. Such constructions are attractive due to their small storage requirement as well as fast generation capability \cite{Liu16}. For given flock size $N$, we define $r\triangleq N/M$ as the performance ratio which is upper bounded by one.  In the case of power-of-two lengths, $r=1$ can be achieved by CCC; For non-power-of-two lengths, it is intriguing to know the maximum achievable $r$. By constructing MOCSs with lengths $2^\theta+2^\gamma$ (where $\theta\neq \gamma$ are positive integers)\footnote{See Corollary 5 in Section III for details.}, it is found that $r=1/2$ is achievable.

The remainder of the paper is organized as follow. In Section \ref{Preliminaries and Definitions}, we give the definitions and some notations that will be used throughout this paper. Then we present the several constructions of GCSs and MOCSs with flexible lengths based on GBFs in Section \ref{Constructions of MOCS}. Finally, we summarize our paper in Section \ref{Conclusion}.

\section{Preliminaries and Definitions}\label{Preliminaries and Definitions}
\subsection{Correlation Functions}
For two $L$-length sequences ${\bm c}=(c_0,c_1,\ldots,c_{L-1})$ and ${\bm d}=(d_0,d_1,\ldots,d_{L-1})$ over ${\mathbb Z}_q = \{0,1,\ldots,q-1\}$. The {\em aperiodic cross-correlation function} of ${\bm c}$ and  ${\bm d}$ at time-shift $u$ is defined as
	\begin{equation*}
 	\rho({\bm c},{\bm d};u)=
    \begin{cases}
	\sum\limits_{k=0}^{L-1-u}\xi^{c_{k+u}-d_k},   \qquad  0 \leq u\leq L-1\\
	\sum\limits_{k=0}^{L-1+u}\xi^{c_{k}-d_{k-u}}, \qquad  -L+1 \leq u<0
	\end{cases}
  	\label{eq:cross}
	\end{equation*}
where $\xi=e^{2\pi \sqrt{-1}/q}$. When ${\bm d}={\bm c}$, the correlation $\rho({\bm c};u) \triangleq \rho({\bm c},{\bm c};u)$ refers to the {\em aperiodic autocorrelation function} of ${\bm c}$.

\subsection{Golay Complementary Set and Mutually Orthogonal Complementary Set}
Consider a set of $M$ sequence sets ${\cal C}=\{C^{p},p=0,1,\ldots,M-1\}$ where each sequence set $C^{p}$ contains $N$ sequences $C^{p}=\{{\bm c}^{p}_{0}, {\bm c}^{p}_{1}, \ldots, {\bm c}^{p}_{N-1}\}$ and each sequence has length $L$.
\begin{defin}
The set ${\cal C}$ is called a {\em mutually orthogonal complementary set} (MOCS), denoted by $(M,N,L)$-MOCS, if
    \begin{align} \label{eq:CCC}
	\lefteqn{{\rho}(C^{p_{1}},C^{p_{2}};u)=\sum_{n=0}^{N-1}\rho({\bm c}_{n}^{p_{1}}, {\bm c}_{n}^{p_{2}};u)} \nn \\
	& =
	\left\{\begin{array}{ll}
	NL,& u=0,~p_{1}=p_{2} \\
	0,& 0 <|u| < L,~p_{1}=p_{2} \\
	0,& |u| < L,~p_{1} \neq p_{2} \\
	\end{array}\right.
    \end{align}
where $M$ is the set size, $N$ the flock size, and $L$ the sequence length. In the above definition, each sequence set $C^{p}$ is called a {\em Golay Complementary Set}, denoted by $(N,L)$-GCS. In addition, any two distinctive GCSs in the set ${\cal C}$ are said to be {\em mutually orthogonal}.
\end{defin}
\begin{lemma}\label{lemma:CCC_bound}
For a $(M,N,L)$-MOCS, the upper bound on set size satisfies the inequality
\begin{equation}\label{CCC_bound}
M \leq N.
\end{equation}
When $M=N$, it is called a {\em Complete Complementary Code} (CCC) \cite{Han_11}.
\end{lemma}

\subsection{Generalized Boolean Functions (GBFs)}
A GBF $f$ is a function of $m$ variables $x_{1},x_{2},\ldots,x_{m}$ which maps any binary vector $(x_1,x_2,\ldots,x_m)$ in $\mathbb{Z}_2^m$ to a value in $\mathbb{Z}_q$ \cite{Paterson00}. We define a product of $r$ variables, $x_{i_1}x_{i_2}\cdots x_{i_r}$, to be a Boolean monomial of degree $r$. For example, $x_1x_2x_3$ is a monomial of degree 3. We list all the monomials as follows:
\[
1,x_{1},x_{2},\ldots,x_{m},x_{1}x_{2},x_{1}x_{3},\ldots,x_{m-1}x_{m},\ldots,x_{1}x_{2}\cdots x_{m}.
\]
For $q=2$, any Boolean function $f$ can be uniquely expressed as a linear combination of these $2^{m}$ monomials, where the coefficient of each monomial belongs to $\mathbb{Z}_2$ \cite{Sloane}. The resulting expression for $f$ is called the {\em algebraic normal form}. Then, we can define a $2^m$-length sequence ${\bm f}=(f_0,f_1,\ldots,f_{2^m-1})$ for any GBF $f$. The $i$th element of $\bm f$ is given by $f_i = f(i_1,i_2,\dots, i_m)$ where $(i_1,i_2,\ldots,i_m)$ is the binary representation of the integer $i=\sum_{j=1}^{m}i_{j}2^{j-1}$.  For example, for $f=x_{1}x_{2}+3$ with $m=3$ and $q=4$, we have the sequence ${\bm f}=(3,3,3,0,3,3,3,0)$.

For a GBF $f$ with $m$ variables, the corresponding sequence ${\bm f}$ defined above is of length $2^{m}$. Considering the sequence of length $\neq 2^{m}$, we define the corresponding truncated sequence ${\bm f}^{(L)}$ of the GBF $f$ by removing the last $2^{m}-L$ elements of the sequence ${\bm f}$ \cite{Super_16}. That is, ${\bm f}^{(L)}=(f_0,f_1,\ldots,f_{L-1})$ is a sequence of length $L$ with $f_{i}=f(i_{1},i_{2},\ldots,i_{m})$ for $i = 0,1,\ldots, L-1$ and $(i_{1},i_{2},\ldots,i_{m})$ is the binary representation of $i$. For example, if taking $m=3$, and GBF $f=x_{1}x_{2}+1$, we have the corresponding truncated sequence ${\bm f}^{(6)}=(1,1,1,0,1,1)$ and ${\bm f}^{(7)}=(1,1,1,0,1,1,1)$. For the sake of simplicity, the superscript of ${\bm f}^{(L)}$ is disregarded when the length $L$ can be determined from the context.

\section{Constructions of MOCSs with Flexible Lengths}\label{Constructions of MOCS}
In this section, we first provide a construction of GCSs based on GBF. Then we extend the construction to MOCSs of non-power-of-two lengths.
\begin{thm}\label{thm:length-GCS}
For any positive integers $m$ and $k$ where $m\geq2$ and $k \leq m$, we consider a partition of $\{1,2,\ldots,m-1\}$ denoted by $I_{1}, I_{2}, \ldots, I_{k}$. Let $\pi_{\alpha}$ be a bijection from $\mathbb{N}_{m_{\alpha}}=\{1,2,\ldots, m_\alpha\}$ to $I_{\alpha}$ where $m_{\alpha}$ is the order of $I_{\alpha}$ for $\alpha=1,2, \ldots, k$. Let the GBF
 \begin{eqnarray}
 f & = & \frac{q}{2}\sum_{\alpha=1}^{k}\sum_{\beta=1}^{m_\alpha-1}x_{\pi_\alpha(\beta)}x_{\pi_\alpha(\beta+1)}
         + \sum_{l=1}^{m}g_l x_l+g_{0},
         \label{eq:f_form}
 \end{eqnarray}
where $q$ is an even integer and $g_l\in \mathbb{Z}_q$ for $l=0,1,\ldots,m$. For a positive integer $t$ with $0 \leq t\leq m-1$, if $t\neq 0$, then there exist integers $\alpha^{\prime}$ and $\beta^{\prime}$ such that $t =\sum_{\alpha=1}^{\alpha^{\prime}-1}m_{\alpha}+\beta^{\prime}$ where $1\leq \beta^{\prime}\leq m_{\alpha^{\prime}}$. Note that if $t\leq m_{1}$, we let $\alpha^{\prime}=1$ and $\beta^{\prime}=t$. When $t \neq 0$, we need an additional condition below.
\begin{eqnarray}\label{condition}
\{\pi_{1}(1),\pi_{1}(2),\ldots,\pi_{1}(m_{1}),\pi_{2}(1),\ldots,\pi_{2}(m_{2}),\nn \\
\pi_{3}(1),\ldots,
\pi_{\alpha^{\prime}}(1),\ldots,\pi_{\alpha^{\prime}}(\beta^{\prime})\}=\{1,2,\ldots,t\}.
\end{eqnarray}
Then, $C=\left\{{\bm c}_{0}, {\bm c}_{1}, \ldots, {\bm c}_{2^{k+1}-1}\right\}$ is a $(2^{k+1},2^{m-1}+2^{t})$-GCS where
\begin{eqnarray*}
 {\bm c}_{n} & = & {\bm f}+\frac{q}{2}\left(\sum_{\alpha=1}^{k}n_{\alpha}{\bm x}_{\pi_{\alpha}(1)}+n_{k+1}{\bm x}_{m}\right),
\end{eqnarray*}
and $(n_{1}, n_{2}, \ldots, n_{k+1})$ is the binary representations of $n$ for $n=0,1, \ldots, 2^{k+1}-1$.
\end{thm}
\begin{IEEEproof}
The proof is given in Appendix \ref{apxB}.
\end{IEEEproof}

Compared with \cite{Super_16} and \cite{Super_172}, the proposed GCSs can exist more lengths by setting a generalized variable $t$ in Theorem \ref{thm:length-GCS}. In addition, the constructed GCSs from Theorem \ref{thm:length-GCS} can include the results in \cite{Super_16,Super_172}. For example, Theorem \ref{thm:length-GCS} includes \cite[Th. 6]{Super_16} as a special case when $t=0$. In addition, if we let $0 \leq t\leq m_{1}$, Theorem \ref{thm:length-GCS} can be reduced to \cite{Super_172}. Then we will illustrate the relationship between Theorem \ref{thm:length-GCS} and the results in \cite{Super_16,Super_172} via an example.

In the following example and from here onwards, we let $\pi(a_1,a_2,\cdots,a_k)$ represent a mapping such that $\pi(1)=a_1$, $\pi(2)=a_2$,$\cdots$, and $\pi(k)=a_k$.
\begin{eg}\label{GCSex1}
Taking $q=4$, $m=6$, and $k=2$, we let $I_{1}=\{1,2\}$, $I_{2}=\{3,4,5\}$, $\pi_{1}=(1,2)$, $\pi_{2}=(3,4,5)$, and then the Boolean function is $f=2x_1x_2+2x_3x_4+2x_4x_5$ by letting all $g_l$'s$=0$ in (\ref{eq:f_form}). Let us consider the case $t=2$, we have $\alpha^{\prime}=1$ and $\beta^{\prime}=2$. Since $\{\pi_{1}(1),\pi_{1}(2)\}=\{1,2\}$ satisfying (\ref{condition}), the set $C=\{{\bm f}+2n_{1}{\bm x}_1+2n_{2}{\bm x}_3+2n_{3}{\bm x}_6 : n_{i}\in \mathbb{Z}_2\}$ is an $(8,36)$-GCS from Theorem \ref{thm:length-GCS}. The set $C$ can also be constructed by \cite{Super_172}. Next, if we take $t=3$, $t$ can be expressed as $t=m_1+1$. Hence, we have $\alpha^{\prime}=2$ and $\beta^{\prime}=1$. The condition (\ref{condition}) still holds since $\{\pi_{1}(1),\pi_{1}(2),\pi_{2}(1)\}=\{1,2,3\}$. Therefore, we can obtain another $(8,40)$-GCS from Theorem \ref{thm:length-GCS}. This $(8,40)$-GCS is listed as follows:
\begin{align} \label{GCSC}
 C     & =  \{{\bm c}_{0},{\bm c}_{1},{\bm c}_{2},{\bm c}_{3},
               {\bm c}_{4},{\bm c}_{5},{\bm c}_{6},{\bm c}_{7}\} \\ \nn
       & = \{(0002000200022220000200022220000200020002),\\ \nn
       &  ~~~~~ (0200020002002022020002002022020002000200),\\ \nn
       &  ~~~~~ (0002222000020002000222202220222000022220),\\ \nn
       &  ~~~~~ (0200202202000200020020222022202202002022),\\ \nn
       &  ~~~~~ (0002000200022220000200022220000222202220),\\ \nn
       &  ~~~~~ (0200020002002022020002002022020020222022),\\ \nn
       &  ~~~~~ (0002222000020002000222202220222022200002),\\ \nn
       &  ~~~~~ (0200202202000200020020222022202220220200)\}.
\end{align}
In this case, \cite{Super_172} cannot construct this $(8,40)$-GCS. Furthermore, if taking $t=0$, \cite[Th. 6]{Super_16}, \cite[Th. 4]{Super_172} and Theorem \ref{thm:length-GCS} all give the same $(8,33)$-GCS.
\end{eg}

Next, we construct multiple GCSs based on Theorem \ref{thm:length-GCS} with mutual orthogonal cross-correlation property, that is, a MOCS.
\begin{thm}\label{thm:length-MOCS}
We follow the same notations given in Theorem \ref{thm:length-GCS} with $f$ defined in (\ref{eq:f_form}). For any positive integer $t$ with $m_{1}\leq t \leq m-1$, we can obtain integers $\alpha^{\prime}$ and $\beta^{\prime}$ such that $t =\sum_{\alpha=1}^{\alpha^{\prime}-1}m_{\alpha}+\beta^{\prime}$ where $\alpha^{\prime}\geq 1$ and $1\leq \beta^{\prime} \leq m_{\alpha^{\prime}}$. Note that $\alpha^{\prime}=1$ and $\beta^{\prime}=m_1$ if $t=m_1$. Let $k^{\prime}$ be the largest integer satisfying $\sum_{\alpha=1}^{k^{\prime}}m_{\alpha}\leq t$. If
\begin{eqnarray}\label{thm:length-MOCS_condition}
\{\pi_{1}(1),\pi_{1}(2),\ldots,\pi_{1}(m_{1}),\pi_{2}(1),\ldots,\pi_{2}(m_{2}), \nn\\
\pi_{3}(1),\ldots,
\pi_{\alpha^{\prime}}(1),\ldots,\pi_{\alpha^{\prime}}(\beta^{\prime})\}=\{1,2,\ldots,t\},
\end{eqnarray}
then the set ${\cal C}=\left\{C^{0},C^{1},\cdots,C^{2^{k^{\prime}}-1}\right\}$ where $C^{p} =\left\{{\bm c}_{0}^{p}, {\bm c}_{1}^{p}, \ldots, {\bm c}_{2^{k+1}-1}^{p}\right\}$,
\begin{eqnarray*}
 {\bm c}_{n}^{p}  =  {\bm f}+\frac{q}{2}\left(\sum_{\alpha=1}^{k}n_{\alpha}{\bm x}_{\pi_{\alpha}(1)}+n_{k+1}{\bm x}_{m}
                        +\sum_{\alpha=1}^{k^{\prime}}p_{\alpha}{\bm x}_{\pi_{\alpha}(m_{\alpha})}\right),
\end{eqnarray*}
for $n=0,1, \ldots, 2^{k+1}-1$ and $p=0,1, \ldots, 2^{k^{\prime}}-1$, $(n_{1}, n_{2}, \ldots, n_{k+1})$ and $(p_{1},p_{2},\ldots,p_{k^{\prime}})$ are the binary representations of $n$ and $p$, respectively, is a $(2^{k^{\prime}},2^{k+1},2^{m-1}+2^{t})$-MOCS.

\end{thm}
\begin{IEEEproof}
The proof is given in Appendix \ref{apxC}.
\end{IEEEproof}

Please note that $k^{\prime}=\alpha^{\prime}$ if $\beta^{\prime}=m_{\alpha^{\prime}}$ and $k^{\prime}=\alpha^{\prime}-1$ if $\beta^{\prime}<m_{\alpha^{\prime}}$.

\begin{eg}\label{CCCex1}
Following the same parameters $q,m,k$, permutations $\pi_{1},\pi_{2}$ and Boolean function $f$ given in Example \ref{GCSex1}, we obtain an $(8,40)$-GCS $C^{0}=C$ in (\ref{GCSC}). According to Theorem \ref{thm:length-MOCS}, we have $k'=1$ since $m_1=2\leq t$ and $m_1+m_2=2+3>t$ when $t$ is taken as 3. As a result, we can obtain a $(2,8,40)$-MOCS from Theorem \ref{thm:length-MOCS}. In addition to $C^{0}$, another constituent GCS is $C'=\{{\bm f}+2n_1{\bm x}_1+2n_2{\bm x}_3+2n_3{\bm x}_6+2{\bm x}_2 : n_{i}\in \mathbb{Z}_2\}$ and given by
\begin{align*}
 C^{1} & =  \{{\bm c}^{1}_{0},{\bm c}^{1}_{1},{\bm c}^{1}_{2},{\bm c}^{1}_{3},
               {\bm c}^{1}_{4},{\bm c}^{1}_{5},{\bm c}^{1}_{6},{\bm c}^{1}_{7}\}\\
       & = \{(0002000200022220000200022220000200020002),\\
       &  ~~~~~ (0200020002002022020002002022020002000200),\\
       &  ~~~~~ (0002222000020002000222202220222000022220),\\
       &  ~~~~~ (0200202202000200020020222022202202002022),\\
       &  ~~~~~ (0002000200022220000200022220000222202220),\\
       &  ~~~~~ (0200020002002022020002002022020020222022),\\
       &  ~~~~~ (0002222000020002000222202220222022200002),\\
       &  ~~~~~ (0200202202000200020020222022202220220200)\}.
\end{align*}
Similarly, for $t=4$, we can obtain a $(2,8,48)$-MOCS.
\end{eg}

From Theorem \ref{thm:length-MOCS}, we can obtain a $(2^{k^{\prime}},2^{k+1},2^{m-1}+2^{t})$-MOCS where $m_{1}\leq t \leq m-1$. Additionally taking $t=m-1$, the constructed MOCSs are reduced to CCCs of length $2^{m}$. Theorem \ref{thm:length-MOCS} provides MOCSs of more lengths. Next, we discuss the set size of the constructed MOCSs with length $2^{m-1}+2^{t}$. Let $M_o$ denote the optimal set size of the constructed MOCS, i.e., $M_o=2^{k+1}$. For $m_{k-1}\leq t < m-1$, we only have set size $2^{k^{\prime}}=2^{\alpha^{\prime}-1}=2^{k-1}$. This means that the set size is $2^{k-1}/M_o=1/4$ which is one fourth of the optimal set size. Therefore, we propose a new corollary for expanding the set size of MOCSs.
\begin{crly}\label{thm:length-MOCS2}
We follow the same notations given in Theorem \ref{thm:length-GCS}. If $t =\sum_{\alpha=1}^{k-1}m_{\alpha}+\beta^{\prime}$ for some positive integer $\beta^{\prime} \leq m_{k}$, we let
\[{\bm y}={\bm x}_{m}{\bm x}_{\pi_{k}(\beta^{\prime})}+({\bm 1}\oplus{\bm x}_{m}){\bm x}_{\pi_{k}(m_{k})}.\]
Then, ${\cal C}=\left\{C^{0},C^{1},\cdots,C^{2^{k}-1}\right\}$ is a $(2^{k},2^{k+1},2^{m-1}+2^{t})$-MOCS where $C^{p} =\left\{{\bm c}_{0}^{p}, {\bm c}_{1}^{p}, \ldots, {\bm c}_{2^{k+1}-1}^{p}\right\}$ and
\begin{align*}
 \lefteqn{{\bm c}_{n}^{p}=  {\bm f}} \\
 & +\frac{q}{2}\left(\sum_{\alpha=1}^{k}n_{\alpha}{\bm x}_{\pi_{\alpha}(1)}+n_{k+1}{\bm x}_{m}
                        +\sum_{\alpha=1}^{k-1}p_{\alpha}{\bm x}_{\pi_{\alpha}(m_{\alpha})}+p_{k}{\bm y}\right),
\end{align*}
for $n=0,1, \ldots, 2^{k+1}-1$ and $p=0,1, \ldots, 2^{k}-1$. Note that $(n_{1}, n_{2}, \ldots, n_{k+1})$ and $(p_{1},p_{2},\ldots,p_{k})$ are the binary representations of $n$ and $p$, respectively.
\end{crly}
\begin{IEEEproof}
The proof is given in Appendix \ref{apxD}.
\end{IEEEproof}

From Corollary \ref{thm:length-MOCS2}, the set size of MOCSs is $2^{k}$. This means that the ratio of the set size over the flock size is $M/N=1/2$.

Table \ref{tableChen} shows the existences of MOCSs of set size 4, 8 and 16 for various lengths. Compared with CCCs from \cite{Chen08}, Theorem \ref{thm:length-MOCS} and Corollary \ref{thm:length-MOCS2} can construct MOCSs of non-power-of-two lengths. For example, it can be observed that the MOCSs of set size 4 are available for lengths 18, 20, 24 between $2^4$ and $2^5$.
\begin{table*}[ht]
\caption{The existence of MOCS for various lengths}\label{tableChen}
\centering
\begin{tabular}{|c||c|c|c|c|c|c|c|c|c|c|c|}
\hline
\backslashbox{Set size}{Length} & 4 & 8 & 12 & 16 & 20 & 24 & 32 & 36 & 40 & 48 & 64 \\ \hline
4 \cite{Chen08} & $\surd$  & $\surd$ & -- & $\surd$ & -- & --  & $\surd$ & --  & -- & -- & $\surd$ \\ \hline
4 (Th. \ref{thm:length-MOCS} and Cor. \ref{thm:length-MOCS2}) & -- & $\surd$ & $\surd$ & $\surd$ & $\surd$ & $\surd$ & $\surd$ & $\surd$ & $\surd$ & $\surd$ & $\surd$ \\ \hline \hline

8 \cite{Chen08} & -- & $\surd$ & -- & $\surd$ & -- & -- & $\surd$ & -- & -- & -- & $\surd$ \\ \hline
8 (Th. \ref{thm:length-MOCS} and Cor. \ref{thm:length-MOCS2}) & -- & -- & -- & $\surd$ & -- & $\surd$ & $\surd$ & -- & $\surd$ & $\surd$ & $\surd$ \\ \hline \hline
16 \cite{Chen08} & -- & -- & -- & $\surd$ & -- & -- & $\surd$ & -- & -- & -- & $\surd$ \\ \hline
16 (Th. \ref{thm:length-MOCS} and Cor. \ref{thm:length-MOCS2}) & -- & -- & -- & --  & -- & -- & $\surd$ & -- & -- & $\surd$ & $\surd$ \\ \hline
\end{tabular}
\end{table*}

\begin{eg}\label{CCC2ex1}
Let us consider the same parameter setting in Example \ref{CCCex1}. Following the construction in Corollary \ref{thm:length-MOCS2}, let $C^{2}=\{{\bm f}+2n_1{\bm x}_1+2n_2{\bm x}_3+2n_3{\bm x}_6+2{\bm y} : n_{i}\in \mathbb{Z}_2\}$ and $C^{3}=\{{\bm f}+2n_1{\bm x}_1+2n_2{\bm x}_3+2n_3{\bm x}_6+2{\bm x}_2+2{\bm y} : n_{i}\in \mathbb{Z}_2\}$ where ${\bm y}={\bm x}_6{\bm x}_{\pi_{2}(1)}+({\bm 1}\oplus {\bm x}_6){\bm x}_{\pi_{2}(3)}={\bm x}_6{\bm x}_{3}+({\bm 1}\oplus {\bm x}_6){\bm x}_{5}$. Then $\{C^{0},C^{1},C^{2},C^{3}\}$ forms a $(4,8,48)$-MOCS.
\end{eg}

\begin{eg}\label{CCC2ex2}
Taking $q=2$, $m=6$ and $k=3$, we let $I_{1}= \{1\}$, $I_{2}=\{2,4\}$, $I_{3}=\{3,5\}$, $\pi_{1}=(1)$, $\pi_{2}=(4,2)$, $\pi_{3}=(3,5)$, and then the Boolean function is $f=x_4x_2+x_3x_5$ by letting $g_l=0$ for all $l$ in (\ref{eq:f_form}). Let us consider the case $t=4$, we have $\alpha^{\prime}=3$ and $\beta^{\prime}=1$ since $t=4=m_1+m_2+1$. Due to $\{\pi_{1}(1),\pi_{2}(1),\pi_{2}(2),\pi_{3}(1)\}=\{1,4,2,3\}$, the set ${\cal C}=\{C^{0},C^{1},\cdots,C^{7}\}$ is an $(8,16,48)$-MOCS from Corollary \ref{thm:length-MOCS2} where $C^{p}=\{{\bm f}+n_1{\bm x}_1+n_2{\bm x}_4+n_3{\bm x}_3+n_4{\bm x}_6+p_1{\bm x}_1+p_2{\bm x}_2+p_3({\bm x}_6{\bm x}_3+({\bm 1}\oplus {\bm x}_6){\bm x}_5) : n_i \in \mathbb{Z}_2\}$ and $(p_1,p_2,p_3)$ is the binary representation of $p$.
\end{eg}

In Table \ref{tableCCC3}, MOCSs of lengths between 4 and 40 are listed. Theorem \ref{thm:length-MOCS} and Corollary \ref{thm:length-MOCS2} can construct MOCSs of non-power-of-two lengths but the set sizes can not be equal to the individual flock sizes. The ratio of the set size over the flock size is $M/N=1/2$ for the MOCSs obtained by Corollary \ref{thm:length-MOCS2}.

\begin{table*}[ht]
\tiny
\caption{The existence of Even-length $(M,N)$-MOCS}\label{tableCCC3}
\centering
\begin{tabular}{|c||c|c|c|c|c|c|c|c|c|c|c|c|c|c|}
\hline
\diagbox[width=2.5cm]{{\footnotesize Construction}}{\footnotesize $(M,N)$}{{\footnotesize Length}} &\footnotesize 4 &\footnotesize 6 &\footnotesize 8 &\footnotesize 10 &\footnotesize 12 &\footnotesize 16 &\footnotesize 18 &\footnotesize 20 &\footnotesize 24 &\footnotesize 32 &\footnotesize 34 &\footnotesize 36 &\footnotesize 40 &\footnotesize $M/N$\\
\hline
~\tiny & $(4,4)$ & -- & $(4,4)$ & --  & -- & $(4,4)$ & -- & -- & -- & $(4,4)$ & -- & --  & -- & 1\\
\cline{2-15}
{\footnotesize \cite{MutualGCS_2008}}& -- & -- & $(8,8)$ & -- & -- & $(8,8)$ & -- & -- & -- & $(8,8)$ & -- & -- & -- & 1\\
\cline{2-15}
{\footnotesize \cite{Chen08}}& --  & -- & -- & -- & -- & $(16,16)$ & -- & -- & -- & $(16,16)$ & -- & --  & -- & 1\\
\cline{2-15}
~ & -- & -- & -- & -- & -- & -- & -- & -- & --  & $(32,32)$ & -- & --  & -- & 1\\
\hline
\hline
~ & $(2,4)$ & -- & $(2,4)$ & -- & -- & $(2,4)$ & -- & -- & -- & $(2,4)$ & -- & -- & -- & $1/2$\\
\cline{2-15}
~ & -- & $(2,8)$ & -- & $(2,8)$ & $(2,8)$ & -- & $(2,8)$ & $(2,8)$ &$(2,8)$ & -- & $(2,8)$ & $(2,8)$ & $(2,8)$ & $1/4$\\
\cline{2-15}
~ & -- & -- & -- & $(2,16)$ & -- & -- & $(2,16)$ & $(2,16)$ & -- & --  & $(2,16)$ & $(2,16)$ & $(2,16)$ & $1/8$\\
\cline{2-15}
~ & -- & -- & -- & -- &-- & -- & $(2,32)$ & --  & -- & -- & $(2,32)$ & $(2,32)$ & -- & $1/16$\\
\cline{2-15}
~ & -- & -- & -- & -- &-- & -- & -- & --  & -- & -- & $(2,64)$ & --  & -- & $1/32$\\
\cline{2-15}
~ & -- & -- & $(4,8)$ & -- & -- & $(4,8)$ & -- & --  & -- & $(4,8)$ & -- & --  & -- & $1/2$\\
\cline{2-15}
{\footnotesize Theorem \ref{thm:length-MOCS}}& -- & -- & -- & -- & $(4,16)$ & -- & -- & $(4,16)$ & $(4,16)$ & -- & -- & $(4,16)$ & $(4,16)$ & $1/4$\\ \cline{2-15}
~ & -- & -- & -- & -- & -- & -- & -- & $(4,32)$ & -- & -- & --& $(4,32)$ & $(4,32)$ & $1/8$\\
\cline{2-15}
~ & -- & -- & -- & -- & -- & -- & --& -- & -- & -- & -- & $(4,64)$ & -- & $1/16$\\
\cline{2-15}
~ & -- & -- & -- & -- & -- & $(8,16)$ & -- & --  & -- & $(8,16)$ & -- & -- & --& $1/2$ \\
\cline{2-15}
~ & -- & -- & -- & -- & -- & -- & -- & -- & $(8,32)$ & -- & -- & -- & $(8,32)$ & $1/4$\\
\cline{2-15}
~ & -- & -- & -- & -- & -- & -- & -- & -- & -- & -- & -- & -- & $(8,64)$ & $1/8$\\
\cline{2-15}
~ & -- & -- & -- & -- & -- & -- & -- & --  &-- & $(16,32)$ & -- & --  & -- & $1/2$\\
\hline
\hline
~ & $(2,4)$ & $(2,4)$ & $(2,4)$ & $(2,4)$ & $(2,4)$ & $(2,4)$ & $(2,4)$ & $(2,4)$ & $(2,4)$ & $(2,4)$ & $(2,4)$ & $(2,4)$ & $(2,4)$ & $1/2$\\
\cline{2-15}
~ & -- & -- & $(4,8)$ & -- & $(4,8)$ & $(4,8)$ & -- & $(4,8)$ & $(4,8)$ & $(4,8)$ & -- & $(4,8)$ & $(4,8)$ & $1/2$\\
\cline{2-15}
{\footnotesize Corollary \ref{thm:length-MOCS2}} & -- & -- & -- & -- & -- & $(8,16)$ & -- & --  & $(8,16)$ & $(8,16)$ & -- & -- & $(8,16)$ & $1/2$\\
\cline{2-15}
~ & -- & -- & -- & -- & -- & -- & -- & --  & -- & $(16,32)$ & -- & --  & -- & $1/2$\\
\hline
\end{tabular}
\end{table*}

\section{Conclusion}\label{Conclusion}
In this paper, we have presented new constructions of GCSs and MOCSs based on GBFs with flexible lengths. First, a construction of GCSs with flexible lengths is proposed in Theorem \ref{thm:length-GCS}, which includes the results in \cite{Super_16,Super_172} as special cases. In addition, new constructions of MOCSs have been proposed in Theorem \ref{thm:length-MOCS} and Corollary \ref{thm:length-MOCS2}. The resultant MOCSs can be obtained directly from GBFs without using other tedious sequence operations.

The proposed MOCSs have lengths of the form $2^{m-1}+2^{t}$ where $t>0$. Possible future research includes the study of MOCSs with more available lengths, e.g., $t=0$. In Corollary \ref{thm:length-MOCS2}, the ratio of set size over flock size (i.e., $M/N$) is $1/2$ only. It would be interesting to know that whether CCCs (i.e., $M/N=1$) with non-power-of-two lengths can be constructed by GBFs.

\begin{appendices}
\section{Proof of Theorem \ref{thm:length-GCS}}\label{apxB}
Before starting the proof of Theorem \ref{thm:length-GCS}, we introduce several lemmas which can be regarded as crucial tools to prove our main theorems. We follow the same notations as given in Theorem \ref{thm:length-GCS} and define the common notations as follows. Let two nonnegative integers $i,j<2^m$ have binary representations $(i_1,i_2,\ldots,i_m)$ and $(j_1,j_2,\ldots,j_m)$, respectively.

\begin{lemma}\label{lemma1}\cite[Lemma 2]{Super_16}
For two integers $i$ and $j$ with $0\leq i<j<2^m$ and $m\geq 2$, if $i_s=j_s$ for $s=1,2,\ldots, t$ for a positive integer $t$, then $j\geq i+2^t$.
\end{lemma}
\begin{lemma}\label{lemma2}\cite[Lemma 3]{Super_16}
For an integer $i$ with $2^{m-1}\leq i\leq 2^{m-1}+2^t-1$ where $1\leq t\leq m-1$ and $m\geq 2$, if $i'$ is an integer with binary representation $(i_1,i_2,\ldots, i_{r-1}, 1-i_r, i_{r+1}, \ldots,i_m)$ and $r\leq t$, then we have $2^{m-1}\leq i'\leq 2^{m-1}+2^t-1$.
\end{lemma}
\begin{lemma}\label{lemma:GCSCase1,2}\cite[Lemma 4]{Wu_18}
Following the same definitions and notations given in Theorem \ref{thm:length-GCS}. If $i_{\pi_{\alpha}(1)}\neq j_{\pi_{\alpha}(1)}$ for some $\alpha\in\{1,2, \ldots, k\}$, then for any sequences ${\bm c}_{n} = (c_{n,0}, c_{n,1}, \ldots, c_{n,L-1})\in C$, there exists ${\bm c}^{\prime}_{n} = (c^{\prime}_{n,0}, c^{\prime}_{n,1}, \ldots, c^{\prime}_{n,L-1})={\bm c}_{n}+(q/2){\bm x}_{1}\in C$ such that $\xi^{c_{n,j}-c_{n,i}}+\xi^{c^{\prime}_{n,j}+c^{\prime}_{n,i}}=0$. Similarly if $i_{m}\neq j_{m}$, then for any sequence ${\bm c}_{n} \in C$, there exists ${\bm c}^{\prime}_{n}={\bm c}_{n}+(q/2){\bm x}_{m}\in C$ such that $\xi^{c_{n,j}-c_{n,i}}+\xi^{c^{\prime}_{n,j}+c^{\prime}_{n,i}}=0$.
\end{lemma}

\begin{lemma}\label{lemma4}\cite[Lemma 5]{Wu_18}
Suppose $i_{\pi_\alpha(1)}= j_{\pi_\alpha(1)}$ for  $\alpha=1,2,\ldots,k$. Let us consider three conditions:
\begin{enumerate}
  \item [(C1)] $\hat{\alpha}$ is the largest integer satisfying $i_{\pi_{\alpha}(\beta)} = j_{\pi_{\alpha}(\beta)}$ for $\alpha=1,\ldots,\hat{\alpha}-1$ and $\beta=1,\ldots,m_\alpha$.
  \item [(C2)] $\hat{\beta}$ is the smallest integer such that $i_{\pi_{\hat{\alpha}}(\hat{\beta})} \neq j_{\pi_{\hat{\alpha}}(\hat{\beta})}$.
  \item [(C3)] Let $i^{\prime}$ and $j^{\prime}$ be integers which differ from $i$ and $j$, respectively, in only one position $\pi_{\hat{\alpha}}(\hat{\beta}-1)$. That is, $i_{\pi_{\hat{\alpha}}(\hat{\beta}-1)}^{\prime}= 1-i_{\pi_{\hat{\alpha}}(\hat{\beta}-1)}$ and $j_{\pi_{\hat{\alpha}}(\hat{\beta}-1)}^{\prime}= 1-j_{\pi_{\hat{\alpha}}(\hat{\beta}-1)}$.
\end{enumerate}
If the above conditions are all satisfied, we have
\[
f_j-f_i-f_{j'}+f_{i'}\equiv q/2 \pmod{q}.
\]
\end{lemma}

\begin{IEEEproof}[Proof of Theorem \ref{thm:length-GCS}]
Here we are going to prove that $C$ is a GCS of size $2^{k+1}$ and of length $2^{m-1}+2^{t}$. To this end, it is sufficient to show
\begin{eqnarray}
\rho(C;u) & = & \sum_{n=0}^{2^{k+1}-1}\rho({\bm c}_{n};u) \nn\\
& = & \sum_{n=0}^{2^{k+1}-1}\sum_{i=0}^{L-1-u}\xi^{c_{n,i+u}-c_{n,i}}\nn\\
& = & \sum_{i=0}^{L-1-u}\sum_{n=0}^{2^{k+1}-1}\xi^{c_{n,i+u}-c_{n,i}}=0
\label{eq:GCS_Auto}
\end{eqnarray}
for $0<u<2^{m-1}+2^{t}-1$. For any integer $i<2^{m}+2^{t}-1-u$ with binary representation $(i_1,i_2,\ldots,i_m)$, we let $j=i+u$ with binary representation $(j_1,j_2,\ldots,j_m)$. Then, we will show that (\ref{eq:GCS_Auto}) holds by considering the following four cases.

{\it Case 1:}
If $i_{\pi_{\alpha}(1)}\neq j_{\pi_{\alpha}(1)}$ for some $\alpha\in\{1,2, \ldots, k\}$, according to Lemma \ref{lemma:GCSCase1,2}, we can obtain
\[
\sum_{n=0}^{2^{k+1}-1}\xi^{c_{n,j}-c_{n,i}}=0.
\]

{\it Case 2:}
In this case, we assume $i_{\pi_{\alpha}(1)} = j_{\pi_{\alpha}(1)}$ for all $\alpha\in\{1,2, \ldots, k\}$ and $i_{m}=j_{m}=0$. Here, there exist $\hat{\alpha}$ and $\hat{\beta}$ satisfying the conditions (C1) and (C2) in Lemma \ref{lemma4}. Let $i'$ and $j'$ be integers fulfilling condition (C3) in Lemma \ref{lemma4}. According to Lemma \ref{lemma4}, we have $j^{\prime}=i^{\prime}+u$ and $f_j-f_i-f_{j'}+f_{i'}\equiv q/2 \pmod{q}$. Therefore,
\begin{eqnarray*}
c_{n,j}-c_{n,i}-c_{n,j'}+c_{n,i'} & = & f_j-f_i-f_{j'}+f_{i'} \\
  & \equiv &  q/2 \pmod{q}.
\end{eqnarray*}
Then the above equation implies
\[
\xi^{c_{n,j}-c_{n,i}}+\xi^{c_{n,j'}-c_{n,i'}}=0.
\]

{\it Case 3:}
In this case, suppose $i_{\pi_{\alpha}(1)} = j_{\pi_{\alpha}(1)}$ for all $\alpha\in\{1,2, \ldots, k\}$ and $i_{m}=j_{m}=1$ which means $2^{m-1}\leq i,j \leq 2^{m-1}+2^{t}-1$. Here, there exist $\hat{\alpha}$ and $\hat{\beta}$ satisfying the conditions (C1) and (C2) in Lemma \ref{lemma4}. Then, we will have $\pi_{\hat{\alpha}}(\hat{\beta}) \leq t$. Suppose not, we assume $\pi_{\hat{\alpha}}(\hat{\beta}) > t$. Hence, we obtain that $i_{s}=j_{s}$ for $s=1,2,\ldots,t$ since $\{\pi_{1}(1),\ldots,\pi_{1}(m_{1}),\pi_{2}(1),\ldots,\pi_{\alpha^{\prime}}(\beta^{\prime})\}=\{1,2,\ldots,t\}$. According to Lemma \ref{lemma1}, we have $j \geq i+2^{t} \geq 2^{m-1}+2^{t}$ which contradicts the assumption. Thus, we have $\pi_{\hat{\alpha}}(\hat{\beta}) \leq t$. Then let $i^{\prime}$ and $j^{\prime}$ be integers fulfilling condition (C3) in Lemma \ref{lemma4}. Since $\pi_{\hat{\alpha}}(\hat{\beta}) \leq t$, we also have $\pi_{\hat{\alpha}}(\hat{\beta}-1) \leq t$. According to Lemma \ref{lemma2}, we obtain $2^{m-1} \leq i',j' \leq 2^{m-1}+2^{\pi_{\hat{\alpha}}(\hat{\beta}-1)}-1 \leq 2^{m-1}+2^t-1$. Therefore, for any integers $i$ and $j$ with $2^{m-1}\leq i,j \leq 2^{m-1}+2^{t}-1$, there exist integers $i',j'\leq 2^{m-1}+2^{t}-1$ and $j^{\prime}=i^{\prime}+u$. Therefore, we obtain
\[
\xi^{c_{n,j}-c_{n,i}}+\xi^{c_{n,j'}-c_{n,i'}}=0.
\] according to Lemma \ref{lemma4}.

{\it Case 4:}
If $i_{m}\neq j_{m}$, according to Lemma \ref{lemma:GCSCase1,2}, we also have
\[
\sum_{n=0}^{2^{k+1}-1}\xi^{c_{n,j}-c_{n,i}}=0.
\]
According to Case 1 to Case 4, we can confirm that $C$ is a GCS of length $L=2^{m-1}+2^{t}$.
\end{IEEEproof}
\section{Proof of Theorem \ref{thm:length-MOCS}}\label{apxC}
\begin{lemma}\label{lemma:BF_com}
Let ${\bm x}_{n_{1}},{\bm x}_{n_{2}},\ldots,{\bm x}_{n_{k}}$ be the sequences corresponding to Boolean functions ${ x}_{n_{1}}, { x}_{n_{2}},$ $\ldots,{ x}_{n_{k}}$, respectively, where $n_{1}<n_{2}<\cdots<n_{k}$. Let a binary sequence ${\bm d}=(d_{0},d_{1},\ldots,d_{L-1})={\bm x}_{n_{1}}\oplus{\bm x}_{n_{2}}\oplus \cdots \oplus{\bm x}_{n_{k}}$ be the sum of ${\bm x}_{n_{1}},{\bm x}_{n_{2}},\ldots,{\bm x}_{n_{k}}$. If $2^{n_{1}}|L$, we can obtain that the Hamming weight of ${\bm d}$ is $L/2$.
\end{lemma}
\begin{IEEEproof}
For any integers $i$, if $2^{n_{1}}|L$, then there exists an integer $i'$ which is different from $i$ in only one position $n_{1}$, i.e., $(i'_1,i'_2,\ldots,i'_m)=(i_1,i_2,\ldots, i_{n_{1}-1}, 1-i_{n_{1}}, i_{n_{1}+1}, \ldots,i_m)$. Thus,
\begin{align*}
d_{i'}-d_{i} & =  (i'_{n_{1}}\oplus i'_{n_{2}}\oplus \cdots \oplus i'_{n_{k}})-(i_{n_{1}}\oplus i_{n_{2}}\oplus \cdots \oplus i_{n_{k}}) \\
             & \equiv  i'_{n_{1}}-i_{n_{1}} \equiv 1 \pmod{2}.
\end{align*}
Therefore, $d_{i'}=1-d_{i}$ which means there are half 0's and half 1's for the binary sequence ${\bm d}$. That is, the Hamming weight of ${\bm d}$ is $L/2$.
\end{IEEEproof}
\begin{IEEEproof}[Proof of Theorem \ref{thm:length-MOCS}]
We will show that any two distinct sets $C^{e}$ and $C^{p}$ where $0\leq e\neq p\leq 2^{k^{\prime}}-1$ satisfy the ideal cross-correlation property. Let ${\bm c}_{n}^{e}=(c_{n,0}^{e}, c_{n,1}^{e}, \ldots, c_{n,L-1}^{e}) \in C^{e}$ and ${\bm c}_{n}^{p}=(c_{n,0}^{p}, c_{n,1}^{p}, \ldots, c_{n,L-1}^{p}) \in C^{p}$ where $L=2^{m-1}+2^{t}$. For $0\leq u<L-1$, we have to show
\begin{eqnarray}
\rho(C^{e},C^{p};u) & = & \sum_{n=0}^{2^{k+1}-1}\rho({\bm c}_{n}^{e},{\bm c}_{n}^{p};u) \nonumber\\
 & = & \sum_{n=0}^{2^{k+1}-1}\sum_{i=0}^{L-1-u}\xi^{c_{n,i+u}^{e}-c_{n,i}^{p}}\nn\\
 & = & \sum_{i=0}^{L-1-u}\sum_{n=0}^{2^{k+1}-1}\xi^{c_{n,i+u}^{e}-c_{n,i}^{p}}=0.
\label{eq:cross_CCCform}
\end{eqnarray}
For any integer $i<2^{m-1}+2^{t}-1-u$ with binary representation $(i_1,i_2,\ldots,i_m)$, we let $j=i+u$ with binary representation $(j_1,j_2,\ldots,j_m)$. Then, we will show that (\ref{eq:cross_CCCform}) holds by considering the following four cases.

{\it Case 1:}
Suppose $i_{\pi_{\alpha}(1)}\neq j_{\pi_{\alpha}(1)}$ for some $\alpha\in\{1,2, \ldots, k\}$, according to Lemma \ref{lemma:GCSCase1,2}, we can obtain
\[\sum_{n=0}^{2^{k+1}-1}\xi^{c_{n,j}^{e}-c_{n,i}^{p}}=0.\]

{\it Case 2:}
In this case, we have $i_{\pi_{\alpha}(1)} = j_{\pi_{\alpha}(1)}$ for all $\alpha\in\{1,2, \ldots, k\}$ and $i_m=j_m=0$. Let $\hat{\alpha},\hat{\beta}, i^{\prime}$, and $j^{\prime}$ be given as those in Case 2 of the proof of Theorem \ref{thm:length-GCS}. Similarly, according to Lemma \ref{lemma4}, we have $f_j-f_i-f_{j'}+f_{i'}\equiv q/2 \pmod{q}$. Thus,
\[
\xi^{c_{n,j}^{e}-c_{n,i}^{p}}+\xi^{c_{n,j'}^{e}-c_{n,i'}^{p}}=0.
\]

{\it Case 3:}
In this case, we have $i_{\pi_{\alpha}(1)} = j_{\pi_{\alpha}(1)}$ for all $\alpha\in\{1,2, \ldots, k\}$ and $i_m=j_m=1$. Let $\hat{\alpha},\hat{\beta}, i^{\prime}$, and $j^{\prime}$ be given as those in Case 3 of the proof of the Theorem \ref{thm:length-GCS}. According to Lemma \ref{lemma4}, we also have $f_j-f_i-f_{j'}+f_{i'}\equiv q/2 \pmod{q}$. Then,
\[
\xi^{c_{n,j}^{e}-c_{n,i}^{p}}+\xi^{c_{n,j'}^{e}-c_{n,i'}^{p}}=0.
\]

{\it Case 4:}
Suppose $i_{m}\neq j_{m}$, according to Lemma \ref{lemma:GCSCase1,2}, we have
\[\sum_{n=0}^{2^{k+1}-1}\xi^{c_{n,j}^{e}-c_{n,i}^{p}}=0.\]

Combining these four cases, we can obtain that equation (\ref{eq:cross_CCCform}) holds for $0<u<L-1$. Now, it remains to show that for $u=0$,
\begin{eqnarray*}
\rho(C^{e},C^{p};0)& = & \sum_{n=0}^{2^{k+1}-1}\rho({\bm c}_{n}^{e},{\bm c}_{n}^{p};0) \\
 & = & \sum_{n=0}^{2^{k+1}-1}\sum_{i=0}^{L-1}\xi^{c_{n,i}^{e}-c_{n,i}^{p}}=0.
\end{eqnarray*}
For any integer $n<2^{k+1}$, we have
\begin{equation*}
{\bm c}_{n}^{e}-{\bm c}_{n}^{p} \equiv  \frac{q}{2}{\bm d} \pmod{q}
\end{equation*}
where
\begin{eqnarray*}
{\bm d} & = & (e_{1}\oplus p_{1}){\bm x}_{\pi_{1}(m_{1})}\oplus(e_{2}\oplus p_{2}){\bm x}_{\pi_{2}(m_{2})}\oplus\cdots \\
 & & \oplus(e_{k^{\prime}}\oplus p_{k^{\prime}}){\bm x}_{\pi_{k^{\prime}}(m_{k^{\prime}})}
\end{eqnarray*}
where $(e_{1}, e_{2}, \ldots,e_{k^{\prime}})$ and $(p_{1}, p_{2}, \ldots,p_{k^{\prime}})$ are the binary representations of $e$ and $p$, respectively. We have ${\bm d} \neq {\bm 0}$ since $e\neq p$. Let $s$ denote the minimum integer in the set $\{\pi_{1}(m_{1}),\pi_{2}(m_{2}),\ldots,\pi_{k^{\prime}}(m_{k^{\prime}})\}$. Due to the condition (\ref{thm:length-MOCS_condition}), we have $\{\pi_{1}(m_{1}),\pi_{2}(m_{2}),\ldots,\pi_{k^{\prime}}(m_{k^{\prime}})\}\subseteq \{1,2,\ldots,t\}$. Hence, we have $s \leq t$ implying $2^{s}|(2^{m-1}+2^{t})$. According to Lemma \ref{lemma:BF_com}, we can obtain that the Hamming weight of ${\bm d}$ is $L/2$. Hence, for $i=0,1, \ldots, 2^{m-1}+2^{t}-1$, there are $L/2$ pairs $(c_{n,i}^{e}, c_{n,i}^{p})$ such that $\xi^{c_{n,i}^{e}-c_{n,i}^{p}}=\xi^{q/2}=-1$ and $L/2$ pairs $(c_{n,i}^{e}, c_{n,i}^{p})$ such that $\xi^{c_{n,i}^{e}-c_{n,i}^{p}}=\xi^{0}=1$, so we have \[\rho({\bm c}_{n}^{e}, {\bm c}_{n}^{p};0)= \sum_{i=0}^{L-1}\xi^{c_{n,i}^{e}-c_{n,i}^{p}}=0\] which completes the proof.
\end{IEEEproof}
\section{Proof of Corollary \ref{thm:length-MOCS2}}\label{apxD}
Corollary \ref{thm:length-MOCS2} can be proved by following the similar logics in Theorem \ref{thm:length-MOCS}. Therefore, we only need to consider some other cases in this proof. We will show that any two distinct sets $C^{e}$ and $C^{p}$ where $0\leq e\neq p\leq 2^{k}-1$ satisfy the ideal cross-correlation property. For $0<u<L-1$ where $L=2^{m-1}+2^{t}$, we also have to show that the equation (\ref{eq:cross_CCCform}) is satisfied. We also consider four cases as follows.

{\it Case 1:}
In this case, we have $i_{\pi_{\alpha}(1)}\neq j_{\pi_{\alpha}(1)}$ for some $\alpha\in\{1,2, \ldots, k\}$. Same as Case 1 of the proof of Theorem \ref{thm:length-MOCS}, we can obtain
\[\sum_{n=0}^{2^{k+1}-1}\xi^{c_{n,i+u}^{e}-c_{n,i}^{p}}=0.\]

{\it Case 2:}
In this case, we have $i_{\pi_{\alpha}(1)} = j_{\pi_{\alpha}(1)}$ for all $\alpha\in\{1,2, \ldots, k\}$ and $i_m=j_m=0$. Let $\hat{\alpha},\hat{\beta}, i^{\prime}$, and $j^{\prime}$ be given as those in Case 2 of the proof of Theorem \ref{thm:length-GCS}. Here, $y_i=i_{m}i_{\pi_{k}(\beta^{\prime})}+(1\oplus i_{m})i_{\pi_{k}(m_{k})}=i_{\pi_{k}(m_{k})}$. Also, $y_{i^{\prime}}=i^{\prime}_{\pi_{k}(m_{k})}$, $y_{j}=j_{\pi_{k}(m_{k})}$, and $y_{j^{\prime}}=j^{\prime}_{\pi_{k}(m_{k})}$. According to Lemma \ref{lemma4}, we have $f_j-f_i-f_{j'}+f_{i'}\equiv q/2 \pmod{q}$. Then,
\begin{align*}
\lefteqn{c_{n,j}^{e}-c_{n,i}^{p}-c_{n,j^{\prime}}^{e}+c_{n,i^{\prime}}^{p} }\\
& =
\left(f_{j}+\frac{q}{2}\sum_{\alpha=1}^{k-1}e_{\alpha}j_{\pi_{\alpha}(m_{\alpha})}+ \frac{q}{2}e_{k}y_{j} \right)\\
& -
\left(f_{i}+\frac{q}{2}\sum_{\alpha=1}^{k-1}p_{\alpha}i_{\pi_{\alpha}(m_{\alpha})}+
\frac{q}{2}p_{k}y_{i} \right)\\
& -
\left(f_{j^{\prime}}+\frac{q}{2}\sum_{\alpha=1}^{k-1}e_{\alpha}j^{\prime}_{\pi_{\alpha}(m_{\alpha})}+
\frac{q}{2}e_{k}y_{j^{\prime}}\right)\\
& +
\left(f_{i^{\prime}}+\frac{q}{2}\sum_{\alpha=1}^{k-1}p_{\alpha}i^{\prime}_{\pi_{\alpha}(m_{\alpha})}+
\frac{q}{2}p_{k}y_{i^{\prime}}\right)\\
& =
\left(f_{j}+\frac{q}{2}\sum_{\alpha=1}^{k-1}e_{\alpha}j_{\pi_{\alpha}(m_{\alpha})}+ \frac{q}{2}e_{k}j_{\pi_{k}(m_{k})} \right)\\
& -
\left(f_{i}+\frac{q}{2}\sum_{\alpha=1}^{k-1}p_{\alpha}i_{\pi_{\alpha}(m_{\alpha})}+
\frac{q}{2}p_{k}i_{\pi_{k}(m_{k})} \right)\\
& -  \left(f_{j^{\prime}}+\frac{q}{2}\sum_{\alpha=1}^{k-1}e_{\alpha}j^{\prime}_{\pi_{\alpha}(m_{\alpha})}+
\frac{q}{2}e_{k}j^{\prime}_{\pi_{k}(m_{k})}\right)\\
& +  \left(f_{i^{\prime}}+\frac{q}{2}\sum_{\alpha=1}^{k-1}p_{\alpha}i^{\prime}_{\pi_{\alpha}(m_{\alpha})}+
\frac{q}{2}p_{k}i^{\prime}_{\pi_{k}(m_{k})}\right)\\
& =  f_{j}-f_{i}-f_{j^{\prime}}+f_{i^{\prime}}\\
&\equiv \frac{q}{2} \pmod{q}
\end{align*}
since $i_{\pi_{\alpha}(m_{\alpha})}=i^{\prime}_{\pi_{\alpha}(m_{\alpha})}$ and $j_{\pi_{\alpha}(m_{\alpha})}=j^{\prime}_{\pi_{\alpha}(m_{\alpha})}$ for $\alpha=1,2,\ldots,k$. This is because $i$ and $i^{\prime}$ differ in only one position $\pi_{\hat{\alpha}(\hat{\beta}-1)}$ and $\hat{\beta}-1<m_{\hat{\alpha}}$; so do $j$ and $j^{\prime}$.
Hence, we can obtain
\[\xi^{c_{n,i+u}^{e}-c_{n,i}^{p}}+\xi^{c_{n,i'+u}^{e}-c_{n,i'}^{p}}=0.\]

{\it Case 3:}
In this case, we have $i_{\pi_{\alpha}(1)} = j_{\pi_{\alpha}(1)}$ for all $\alpha\in\{1,2, \ldots, k\}$ and $i_m=j_m=1$. Different to Case 2, we have $y_i=i_{\pi_{k}(\beta^{\prime})}$, $y_{i^{\prime}}=i^{\prime}_{\pi_{k}(\beta^{\prime})}$, $y_{j}=j_{\pi_{k}(\beta^{\prime})}$, and $y_{j^{\prime}}=j^{\prime}_{\pi_{k}(\beta^{\prime})}$ instead. Let $\hat{\alpha},\hat{\beta}, i^{\prime}$ and $j^{\prime}$ be given as those in Case 4 of the proof of Theorem \ref{thm:length-GCS}.  According to Lemma \ref{lemma4}, we have $f_j-f_i-f_{j'}+f_{i'}\equiv q/2 \pmod{q}$. Thus,
\begin{align}\label{eq}
\lefteqn{c_{n,j}^{e}-c_{n,i}^{p}-c_{n,j^{\prime}}^{e}+c_{n,i^{\prime}}^{p} } \nn\\
& =  \left(f_{j}+\frac{q}{2}\sum_{\alpha=1}^{k-1}e_{\alpha}j_{\pi_{\alpha}(m_{\alpha})}+
\frac{q}{2}e_{k}j_{\pi_{k}(\beta^{\prime})} \right) \nn \\ \nn
& -
\left(f_{i}+\frac{q}{2}\sum_{\alpha=1}^{k-1}p_{\alpha}i_{\pi_{\alpha}(m_{\alpha})}
+\frac{q}{2}p_{k}i_{\pi_{k}(\beta^{\prime})} \right)\\ \nn
& -  \left(f_{j^{\prime}}+\frac{q}{2}\sum_{\alpha=1}^{k-1}e_{\alpha}j^{\prime}_{\pi_{\alpha}(m_{\alpha})}
+\frac{q}{2}e_{k}j^{\prime}_{\pi_{k}(\beta^{\prime})} \right)\\ \nn
& +  \left(f_{i^{\prime}}+\frac{q}{2}\sum_{\alpha=1}^{k-1}p_{\alpha}i^{\prime}_{\pi_{\alpha}(m_{\alpha})}
+\frac{q}{2}p_{k}i^{\prime}_{\pi_{k}(\beta^{\prime})}\right)\\ \nn
& =  f_{j}-f_{i}-f_{j^{\prime}}+f_{i^{\prime}}\\
& +  \frac{q}{2}e^{\prime}_{\alpha}\left(j_{\pi_{k}(\beta^{\prime})}-j^{\prime}_{\pi_{k}(\beta^{\prime})} \right)- \frac{q}{2}p^{\prime}_{\alpha}\left(i_{\pi_{k}(\beta^{\prime})}-i^{\prime}_{\pi_{k}(\beta^{\prime})} \right)
\end{align}
where $i_{\pi_{\alpha}(m_{\alpha})}=i^{\prime}_{\pi_{\alpha}(m_{\alpha})}$ and $j_{\pi_{\alpha}(m_{\alpha})}=j^{\prime}_{\pi_{\alpha}(m_{\alpha})}$ for $\alpha=1,2,\ldots,k$ as mentioned in Case 2. If $\pi_{k}(\beta^{\prime}) = \pi_{k}(m_{k})$, then (\ref{eq}) can be further expressed as $c_{n,j}^{e}-c_{n,i}^{p}-c_{n,j^{\prime}}^{e}+c_{n,i^{\prime}}^{p}=f_{j}-f_{i}-f_{j^{\prime}}+f_{i^{\prime}}\equiv q/2 \pmod{q}$. In another case, if $\pi_{k}(\beta^{\prime}) \neq \pi_{k}(m_{k})$ which implies $\beta^{\prime}\neq m_{k}$, then we shall have $\pi_{\hat{\alpha}}(\hat{\beta})<\pi_{k}(\beta^{\prime}+1)$. Otherwise, $i_{s}=j_{s}$ for $s=1,2,\ldots,t$ since
\begin{eqnarray*}
\{\pi_{1}(1),\pi_{1}(2),\ldots,\pi_{1}(m_{1}),\pi_{2}(1),\ldots,\pi_{2}(m_{2}),\\
\pi_{3}(1),\ldots,\pi_{k}(1),\ldots,\pi_{k}(\beta^{\prime})\}=\{1,2,\ldots,t\}
\end{eqnarray*}
and $\pi_{k(\beta^{\prime})}<\pi_{\hat{\alpha}}(\hat{\beta})$. According to Lemma \ref{lemma1}, we have $j \geq i+2^{t} \geq 2^{m-1}+2^{t}$ which contradicts the assumption that $j<L=2^{m-1}+2^{t}$. Therefore, $\pi_{\hat{\alpha}}(\hat{\beta}) < \pi_{k}(\beta^{\prime}+1)$ implying $\pi_{\hat{\alpha}}(\hat{\beta}-1) < \pi_{k}(\beta^{\prime})$. Thus, we have $i_{\pi_{k}(\beta^{\prime})}=i^{\prime}_{\pi_{k}(\beta^{\prime})}$ and $j_{\pi_{k}(\beta^{\prime})}=j^{\prime}_{\pi_{k}(\beta^{\prime})}$. So we can obtain
 \begin{eqnarray*}
c_{n,j}^{e}-c_{n,i}^{p}-c_{n,j^{\prime}}^{e}+c_{n,i^{\prime}}^{p} & = & f_{j}-f_{i}-f_{j^{\prime}}+f_{i^{\prime}}\\
&\equiv & \frac{q}{2} \pmod{q}
\end{eqnarray*}
implying
\[\xi^{c_{n,i+u}^{e}-c_{n,i}^{p}}+\xi^{c_{n,i'+u}^{e}-c_{n,i'}^{p}}=0.\]

{\it Case 4:}
In this case, we assume $i_{m}\neq j_{m}$. Same as Case 4 of the proof of Theorem \ref{thm:length-MOCS}, we have
\[\sum_{n=0}^{2^{k+1}-1}\xi^{c_{n,i+u}^{e}-c_{n,i}^{p}}=0.\]

Combining these four cases, we can obtain that equation (\ref{eq:cross_CCCform}) holds for $0<u<L-1$. Similarly, it can also be obtained that equation (\ref{eq:cross_CCCform}) holds for $0<u<-L+1$. Now, it remains to show that
\begin{eqnarray*}
\rho(C^{e},C^{p};0) & = & \sum_{n=0}^{2^{k+1}-1}\rho({\bm c}_{n}^{e},{\bm c}_{n}^{p};0)\\
& = & \sum_{n=0}^{2^{k+1}-1}\sum_{i=0}^{L-1}\xi^{c_{n,i}^{e}-c_{n,i}^{p}}=0.
\end{eqnarray*}
For any integer $n<2^{k+1}$, we have
\begin{equation*}
{\bm c}_{n}^{e}-{\bm c}_{n}^{p} \equiv  \frac{q}{2}{\bm d} \pmod{q}
\end{equation*}
where
\begin{eqnarray*}
{\bm d} & = & (e_{1}\oplus p_{1}){\bm x}_{\pi_{1}(m_{1})}\oplus(e_{2}\oplus p_{2}){\bm x}_{\pi_{2}(m_{2})}\oplus\cdots \\
& & \oplus(e_{k-1}\oplus p_{k-1}){\bm x}_{\pi_{k-1}(m_{k-1})} \\
& & \oplus(e_{k}\oplus p_{k})\left({\bm x}_{m}{\bm x}_{\pi_{k}(\beta^{\prime})}+({\bm 1}\oplus {\bm x}_{m}){\bm x}_{\pi_{k}(m_{k})}\right)
\end{eqnarray*}
where $(e_{1}, e_{2}, \ldots,e_{k})$ and $(p_{1}, p_{2}, \ldots,p_{k})$ are the binary representations of $e$ and $p$, respectively. We have ${\bm d} \neq {\bm 0}$ since $e\neq p$. Here, we consider two cases to show that the Hamming weight of ${{\bm d}}$ is $L/2$.

(i). For $e_{k}\oplus p_{k}=0$, we have
\begin{eqnarray*}
{\bm d} & = & (e_{1}\oplus p_{1}){\bm x}_{\pi_{1}(m_{1})}\oplus(e_{2}\oplus p_{2}){\bm x}_{\pi_{2}(m_{2})}\oplus\cdots\\
& & \oplus(e_{k-1}\oplus p_{k-1}){\bm x}_{\pi_{k-1}(m_{k-1})}.
\end{eqnarray*}
Then, arguing as in Case 4 of the proof of Theorem \ref{thm:length-MOCS}, we obtain the same result that ${{\bm d}}$ has Hamming weight $L/2$.

(ii). For $e_{k}\oplus p_{k}=1$, we have
\begin{eqnarray*}
{\bm d} & = & (e_{1}\oplus p_{1}){\bm x}_{\pi_{1}(m_{1})}\oplus(e_{2}\oplus p_{2}){\bm x}_{\pi_{2}(m_{2})}\oplus\cdots\\
& & \oplus(e_{k-1}\oplus p_{k-1}){\bm x}_{\pi_{k-1}(m_{k-1})}\\
& &\oplus {\bm x}_{m}{\bm x}_{\pi_{k}(\beta^{\prime})}\oplus({\bm 1}\oplus{\bm x}_{m}){\bm x}_{\pi_{k}(m_{k})}.
\end{eqnarray*}
We express the sequence ${{\bm d}}$ into two parts, i.e., ${\bm d}=({{\bm d}_1},{{\bm d}_2})$ where ${\bm d}_1=(d_0,d_1,\ldots,d_{2^{m}-1})$ and ${\bm d}_2=(d_{2^{m}},d_{2^{m}+1},\ldots,d_{2^{m}+2^{t}-1})$. For any positive integer $0\leq i \leq 2^{m}-1$, we have $i_m=0$ implying
\begin{eqnarray*}
{\bm d}_1 & = & (e_{1}\oplus p_{1}){\bm x}_{\pi_{1}(m_{1})}\oplus(e_{2}\oplus p_{2}){\bm x}_{\pi_{2}(m_{2})}\oplus\cdots\\
& & \oplus(e_{\alpha^{\prime}-1}\oplus p_{k-1}){\bm x}_{\pi_{k-1}(m_{k-1})}\oplus{\bm x}_{\pi_{k}(m_{k})}.
\end{eqnarray*}
We have $\pi_{\alpha}(m_{\alpha}) \leq m-1$ for $\alpha=1,2,\ldots,k$, so $2^{\pi_{\alpha}(m_{\alpha})}|2^{m-1}$. According to Lemma \ref{lemma:BF_com}, it can be obtained that the Hamming weight of ${\bm d}_1$ is half length, i.e., $2^{m-2}$.

For any positive integer $i$ with $2^{m} \leq i \leq 2^{m}+2^{t}-1$, we have $i_m=1$. Therefore, ${\bm d}_2$ can be expressed as
\begin{eqnarray*}
{\bm d}_2 & = & (e_{1}\oplus p_{1}){\bm x}_{\pi_{1}(m_{1})}\oplus(e_{2}\oplus p_{2}){\bm x}_{\pi_{2}(m_{2})}\oplus\cdots\\
& & \oplus(e_{k-1}\oplus p_{k-1}){\bm x}_{\pi_{k-1}(m_{k-1})}\oplus {\bm x}_{\pi_{k}(\beta^{\prime})}.
\end{eqnarray*}
Since $\{\pi_{1}(m_{1}),\pi_{2}(m_{2}),\cdots,\pi_{k-1}(m_{k-1}),\pi_{k}(\beta^{\prime})\} \in \{1,2,\ldots,t\}$, we have $\pi_{k}(\beta^{\prime}) \leq t$ and $\pi_{\alpha}(m_{\alpha}) \leq t$ for $\alpha=1,2,\ldots,k-1$. Thus, it can be obtained that the Hamming weight of ${\bm d}_2$ is $2^{t}/2=2^{t-1}$ according to Lemma \ref{lemma:BF_com}.

Hence, the Hamming weight of ${\bm d}$ is $2^{m-2}+2^{t-1}=L/2$. There are $L/2$ pairs $(c_{n,i}^{e}, c_{n,i}^{p})$ such that $\xi^{c_{n,i}^{e}-c_{n,i}^{p}}=\xi^{q/2}=-1$ and $L/2$ pairs $(c_{n,i}^{e}, c_{n,i}^{p})$ such that $\xi^{c_{n,i}^{e}-c_{n,i}^{p}}=\xi^{0}=1$, so we have
\[\rho({\bm c}_{n}^{e}, {\bm c}_{n}^{p};0)= \sum_{i=0}^{2^{m}-1}\xi^{c_{n,i}^{e}-c_{n,i}^{p}}=0\] which completes the proof.

\end{appendices}

\section*{Acknowledgment}

The authors would like to thank the anonymous reviewers and Associate Editor Ryan Gabrys for their valuable comments and suggestions.

\bibliographystyle{IEEEtran}
\bibliography{IEEEabrv,2019_reference}

\begin{thebibliography}{10}
\providecommand{\url}[1]{#1}
\csname url@samestyle\endcsname
\providecommand{\newblock}{\relax}
\providecommand{\bibinfo}[2]{#2}
\providecommand{\BIBentrySTDinterwordspacing}{\spaceskip=0pt\relax}
\providecommand{\BIBentryALTinterwordstretchfactor}{4}
\providecommand{\BIBentryALTinterwordspacing}{\spaceskip=\fontdimen2\font plus
\BIBentryALTinterwordstretchfactor\fontdimen3\font minus
  \fontdimen4\font\relax}
\providecommand{\BIBforeignlanguage}[2]{{%
\expandafter\ifx\csname l@#1\endcsname\relax
\typeout{** WARNING: IEEEtran.bst: No hyphenation pattern has been}%
\typeout{** loaded for the language `#1'. Using the pattern for}%
\typeout{** the default language instead.}%
\else
\language=\csname l@#1\endcsname
\fi
#2}}
\providecommand{\BIBdecl}{\relax}
\BIBdecl

\bibitem{Golay}
M.~J.~E. Golay, ``Complementary series,'' \emph{{IRE} Trans. Inf. Theory}, vol.
  IT-7, pp. 82--87, Apr. 1961.

\bibitem{Golay_sets}
C.-C. Tseng and C.~L. Liu, ``Complementary sets of sequences,'' \emph{{IEEE}
  Trans. Inf. Theory}, vol. IT-18, no.~5, pp. 644--652, Sep. 1972.

\bibitem{N_shift}
N.~Suehiro and M.~Hatori, ``{$N$-shift} cross-orthogonal sequences,''
  \emph{{IEEE} Trans. Inf. Theory}, vol.~34, no.~1, pp. 143--146, Jan. 1988.

\bibitem{Bell_CDMA}
S.-M. Tseng and M.~R. Bell, ``Asynchronous multicarrier {DS-CDMA} using
  mutually orthogonal complementary sets of sequences,'' \emph{{IEEE} Trans.
  Commun.}, vol.~48, pp. 53--59, Jan. 2000.

\bibitem{Chen_CDMA}
H.-H. Chen, J.-F. Yeh, and N.~Suehiro, ``A multicarrier {CDMA} architecture
  based on orthogonal complete complementary codes for new generations of
  wideband wireless communications,'' \emph{{IEEE} Commun. Mag.}, vol.~39, pp.
  126--134, Oct. 2001.

\bibitem{Chen2007book}
H.-H. Chen, \emph{The Next Generation CDMA Technologies}.\hskip 1em plus 0.5em
  minus 0.4em\relax Wiley, 2007.

\bibitem{Liu2014}
Z.~Liu, Y.~L. Guan, and U.~Parampalli, ``New complete complementary codes for
  peak-to-mean power control in multi-carrier {CDMA},'' \emph{{IEEE} Trans.
  Commun.}, vol.~62, no.~3, pp. 1105--1113, Mar. 2014.

\bibitem{Liu2015}
Z.~Liu, Y.~L. Guan, and H.-H. Chen, ``Fractional-delay-resilient receiver
  design for interference-free {MC-CDMA} communications based on complete
  complementary codes,'' \emph{{IEEE} Trans. Wireless Commun.}, vol.~14, no.~3,
  pp. 1226--1236, Mar. 2015.

\bibitem{Wang07}
S.~Wang and A.~Abdi, ``{MIMO ISI channel estimation using uncorrelated Golay
  complementary sets},'' \emph{{IEEE} Trans. Veh. Technol.}, vol.~56, no.~5,
  pp. 3024--3039, Sep. 2007.

\bibitem{Li2010}
S.~F. Li, J.~Chen, and L.~Q. Zhang, ``Optimisation of complete complementary
  codes in {MIMO} radar system,'' \emph{Electron. Lett.}, vol.~46, no.~16, pp.
  1157--1159, Aug. 2010.

\bibitem{Tang2014}
J.~Tang, N.~Zhang, Z.~Ma, and B.~Tang, ``{Construction of Doppler resilient
  complete complementary code in MIMO radar},'' \emph{{IEEE} Trans. Signal
  Process.}, vol.~62, no.~18, pp. 4704--4712, Sep. 2014.

\bibitem{ChenICC08}
C.-Y. Chen, Y.-J. Min, K.-Y. Lu, and {C.-c.\ Chao}, ``Cell search for
  cell-based {OFDM} systems using quasi complete complementary codes,'' in
  \emph{{Proc.\ {IEEE} Int.\ Conf.\ Commun.}}, Beijing, China, May 2008, pp.
  4840--4844.

\bibitem{Kojima2014}
T.~Kojima, T.~Tachikawa, A.~Oizumi, Y.~Yamaguchi, and U.~Parampalli, ``A
  disaster prevention broadcasting based on data hiding scheme using complete
  complementary codes,'' in \emph{Proc. Int. Symp. on Inform. Theory and its
  Applicat.}, Melbourne, Australia, Oct. 2014, pp. 45--49.

\bibitem{Golay_RM}
J.~A. Davis and J.~Jedwab, ``Peak-to-mean power control in {OFDM}, {Golay}
  complementary sequences, and {Reed-Muller} codes,'' \emph{{IEEE} Trans. Inf.
  Theory}, vol.~45, no.~7, pp. 2397--2417, Nov. 1999.

\bibitem{Paterson_00}
K.~G. Paterson, ``Generalized {Reed-Muller} codes and power control in {OFDM}
  modulation,'' \emph{{IEEE} Trans. Inf. Theory}, vol.~46, no.~1, pp. 104--120,
  Jan. 2000.

\bibitem{Li_05}
Y.~Li and W.~B. Chu, ``More {Golay} sequences,'' \emph{{IEEE} Trans. Inf.
  Theory}, vol.~51, no.~3, pp. 1141--1145, Mar. 2005.

\bibitem{even-ZCP}
Z.~Liu, U.~Parampalli, and Y.~L. Guan, ``On even-period binary
  {Z}-complementary pairs with large {ZCZs},'' \emph{{IEEE} Signal Process.
  Lett.}, vol.~21, no.~3, pp. 284--287, Mar. 2014.

\bibitem{odd-ZCP}
------, ``Optimal odd-length binary {Z}-complementary pairs,'' \emph{{IEEE}
  Trans. Inf. Theory}, vol.~60, no.~9, pp. 5768--5781, Sep. 2014.

\bibitem{schmidt}
K.-U. Schmidt, ``Complementary sets, generalized {Reed-Muller} codes, and power
  control for {OFDM},'' \emph{{IEEE} Trans. Inf. Theory}, vol.~53, no.~2, pp.
  808--814, Feb. 2007.

\bibitem{Parker_03}
M.~G. Parker and C.~Tellambura, ``A construction for binary sequence sets with
  low peak-to-average power ratio,'' in \emph{Proc.\ {IEEE} Int.\ Symp.\
  Inform.\ Theory}, Lausanne, Switzerland, Jun. 2003, p. 239.

\bibitem{Tellambura_ICC05}
W.~Chen and C.~Tellambura, ``Identifying a class of multiple shift
  complementary sequences in the second order cosets of the first order
  {Reed-Muller} codes,'' in \emph{Proc.\ {IEEE} Int.\ Conf.\ Commun.}, Seoul,
  Korea, May 2005, pp. 618--621.

\bibitem{ChenAAECC06}
{C.-Y.\ Chen, C.-H.\ Wang, and {C.-c.\ Chao}}, ``Complementary sets and
  {Reed-Muller} codes for peak-to-average power ratio reduction in {OFDM},'' in
  \emph{Proc.\ 16th\ Int.\ Symp.\ AAECC, LNCS 3857}, Las Vegas, NV, Feb. 2006,
  pp. 317--327.

\bibitem{MutualGCS_2008}
A.~Rathinakumar and A.~K. Chaturvedi, ``Complete mutually orthogonal {Golay}
  complementary sets from {Reed-Muller} codes,'' \emph{{IEEE} Trans. Inf.
  Theory}, vol.~54, pp. 1339--1346, Mar. 2008.

\bibitem{Chen08}
C.-Y. Chen, C.-H. Wang, and {C.-c.\ Chao}, ``Complete complementary codes and
  generalized {Reed-Muller} codes,'' \emph{{IEEE} Commun. Lett.}, vol.~12, pp.
  849--851, Nov. 2008.

\bibitem{Super_16}
C.-Y. Chen, ``Complementary sets of non-power-of-two length for peak-to-average
  power ratio reduction in {OFDM},'' \emph{{IEEE} Trans. Inf. Theory}, vol.~62,
  no.~12, pp. 7538--7545, Dec. 2016.

\bibitem{Super_172}
------, ``A new construction of {Golay} complementary sets of non-power-of-two
  length based on {Boolean} functions,'' in \emph{Proc.\ IEEE Wireless Commun.
  and Netw. Conf.}, San Francisco, CA, Mar. 2017, pp. 1--6.

\bibitem{Super_18}
------, ``A novel construction of complementary sets with flexible lengths
  based on {Boolean} functions,'' \emph{{IEEE} Commun. Lett.}, vol.~22, no.~2,
  pp. 260--263, Feb. 2018.

\bibitem{Das_18}
S.~Das, S.~Budi\v{s}in, S.~Majhi, Z.~Liu, and Y.~L. Guan, ``A multiplier-free
  generator for polyphase complete complementary codes,'' \emph{{IEEE} Trans.
  Signal Process.}, vol.~66, no.~5, pp. 1184--1196, Mar. 2018.

\bibitem{Das_18_2}
S.~Das, S.~Majhi, and Z.~Liu, ``A novel class of complete complementary codes
  and their applications for {APU} matrices,'' \emph{{IEEE} Signal Process.
  Lett.}, vol.~25, no.~9, pp. 1300--1304, Sep. 2018.

\bibitem{Das_19}
S.~Das, S.~Majhi, S.~Budi\v{s}in, and Z.~Liu, ``A new construction framework
  for polyphase complete complementary codes with various lengths,''
  \emph{{IEEE} Trans. Signal Process.}, vol.~67, no.~10, pp. 2639--2648, May
  2019.

\bibitem{Liu16}
Z.~Liu and Y.~L. Guan, ``16-{QAM} almost-complementary sequences with low
  {PMEPR},'' \emph{{IEEE} Trans. Commun.}, vol.~64, no.~2, pp. 668--679, Feb.
  2016.

\bibitem{Han_11}
C.~Han, N.~Suehiro, and T.~Hashimoto, ``A systematic framework for the
  construction of optimal complete complementary codes,'' \emph{{IEEE} Trans.
  Inf. Theory}, vol.~57, no.~9, pp. 6033--6042, Sep. 2011.

\bibitem{Paterson00}
K.~G. Paterson, ``Generalized {Reed-Muller} codes and power control in {OFDM}
  modulation,'' \emph{{IEEE} Trans. Inf. Theory}, vol.~46, pp. 104--120, Jan.
  2000.

\bibitem{Sloane}
F.~J. MacWilliams and N.~J.~A. {Sloane}, \emph{The Theory of Error Correcting
  Codes}.\hskip 1em plus 0.5em minus 0.4em\relax Amsterdam, The Netherlands:
  North-Holland, 1977.

\bibitem{Wu_18}
S.-W. Wu and C.-Y. Chen, ``Optimal {Z}-complementary sequence sets with good
  peak-to-average power-ratio property,'' \emph{{IEEE} Signal Process. Lett.},
  vol.~25, no.~10, pp. 1500--1504, Oct. 2018.

\end{thebibliography}

\begin{IEEEbiographynophoto}{Shing-Wei~Wu}
received his B.S. degree in electrical engineering from the Chinese Culture University, Taipei, in 2016 and  M.S. degree in engineering science from the National Cheng Kung University, Tainan, Taiwan in 2018. Currently, he is pursuing the
Ph.D. degree in engineering science from the National Cheng Kung University, Tainan, Taiwan. His research interests include sequence design
and its applications on communications.

\end{IEEEbiographynophoto}

\begin{IEEEbiographynophoto}{Chao-Yu Chen}(M'17)
received the B.S. degree in electrical engineering from the National Tsing Hua
University (NTHU), Hsinchu, in 2000 and the M.S. and Ph.D. degrees in communications engineering from NTHU, Hsinchu, in 2002 and 2009, respectively, under the supervision of Prof. Chi-chao Chao.

He was a visiting Ph.D. student with the University of California, Davis, from 2008 to 2009 (with Prof. Shu Lin).
From 2009 to 2016, he was a technical manager in Communication System Design division, Mediatek Inc., Hsinchu, Taiwan. From July 2018 to August 2018, he was with the University of California, Davis, as a visiting scholar (with Prof. Shu Lin). Since February 2016, he has been a Faculty Member with the National Cheng Kung University, Tainan, Taiwan, where he is currently an assistant professor at the Department of Engineering Science. His current research interests include sequence design, error-correcting codes, digital communications, and wireless networks.

Dr. Chen was a recipient of the 15th Y. Z. Hsu Science Paper Award administered by Far Eastern Y. Z. Hsu Science and Technology Memorial Foundation, Taiwan, in 2017 and the Best Paper Award for Young Scholars by the IEEE Information Theory Society Taipei Chapter and the IEEE Communications Society Taipei/Tainan Chapter in 2018. Since January 2019, he serves as the Vice Chair of the IEEE Information Theory Society Tainan Chapter.
\end{IEEEbiographynophoto}

\begin{IEEEbiographynophoto}{Zilong Liu}
received his PhD (2014) from School of Electrical and Electronic Engineering, Nanyang Technological University (NTU), Master Degree (2007) in the Department of Electronic Engineering from Tsinghua University, and Bachelor Degree (2004) in the School of Electronics and Information Engineering from Huazhong University of Science and Technology (HUST). He is a Lecturer at the School of Computer Science and Electronic Engineering, University of Essex. From Jan. 2018 to Nov. 2019, he was a Senior Research Fellow at the Institute for Communication Systems (ICS), Home of the 5G Innovation Centre (5GIC), University of Surrey. Prior to his career in UK, he spent 9.5 years in NTU, Singapore, first as a Research Associate (Jul. 2008 to Oct. 2014) and then a Research Fellow (Nov. 2014 to Dec. 2018). His research lies in the interplay of communication, coding, signal processing, and a wide range of mathematical tools (e.g., number theory, abstract algebra, and convex optimization). He is extremely in attacking research problems that arise in practical communication systems.  Details of his research can be found at: https://sites.google.com/site/zilongliu2357.
\end{IEEEbiographynophoto}

\end{document}